# A Simple Method for More Precise Pulse-Height Fitting in Sparsley Sampled Data Using Pulse-Shape-Archetype Information, Especially Suited to Ultra-Short-Laser Pulsetrains.

**JINGSEN TANG, NICOLAS SOULELES, ANNA HWANG, LUKE COULTER, DENISA BANI, ROBIN S. MARJORIBANKS.**

*Department of Physics, University of Toronto, 60 Saint George St, Toronto, Ontario, M5S 1A7, Canada*
*jingsen.tang@mail.utoronto.ca

## Abstract

This paper presents a simple optical data reconstruction method designed for sparsely sampled, repetitive ultrafast-laser pulse signals. In such cases, the detector and oscilloscope often limit the waveform, and only the pulse energy or height is sought. When each pulse has a well-characterized and repeatable shape, this information can be used as an archetype to extract amplitude and timing parameters with far greater accuracy than general interpolation or fitting methods. The approach is broadly applicable to ultrafast optical diagnostics such as reflectometry and LiDAR. Demonstrations across different detection systems show consistent performance improvements, with amplitude-retrieval errors reduced by more than a factor of two under undersampled conditions. The method provides a straightforward and robust strategy for enhancing precision in optical waveform measurements.

## 1. INTRODUCTION

Accurate pulse-amplitude (or energy) and relative pulse timing estimation from collected traces is crucial for ultrashort laser research such as reflectometry, light detection and ranging (LiDAR) systems, and optical coherence tomography[1−5]. Photodiode detectors combined with digital oscilloscopes are popular detection systems for time-resolved ultrashort laser pulse measurements. However, analog oscilloscopes were long the standard and still have their place today since they give continuous traces without aliasing.

For digital-scopes, the problem of aliasing of repetitive signals is well recognized, and can lead to artifacts (see Fig. 1). When the peaks of a fast, well-defined, repetitive, waveform are sparsely sampled with a shifting phase relationship from pulse to pulse, the digitized data-points may present falsely as a series of pulses of fluctuating amplitudes, delivering an artifact pattern which may then be accepted at face value, by some unsophisticated users. This aliasing becomes even more pronounced when the effective sampling rate is reduced, such as when multiple channels share the digitizer.

This paper addresses this issue: in many quasi-repetitive signals, such as oscilloscope waveforms of ultrashort laser pulse trains, the canonical shape of pulses in a pulse-train may be well identified, but this extra information is wasted when using a needlessly generalized form of waveform-fitting. We describe a specialized method, archetype parametric fitting (APF), of fitting quasi-repetitive waveforms that uses one pulse as an archetype, and best fits the digitized waveform using the archetype exclusively – like a basis vector. What is extracted are the pulse heights and the timings of the pulses within the pulse-train.

**(a)**

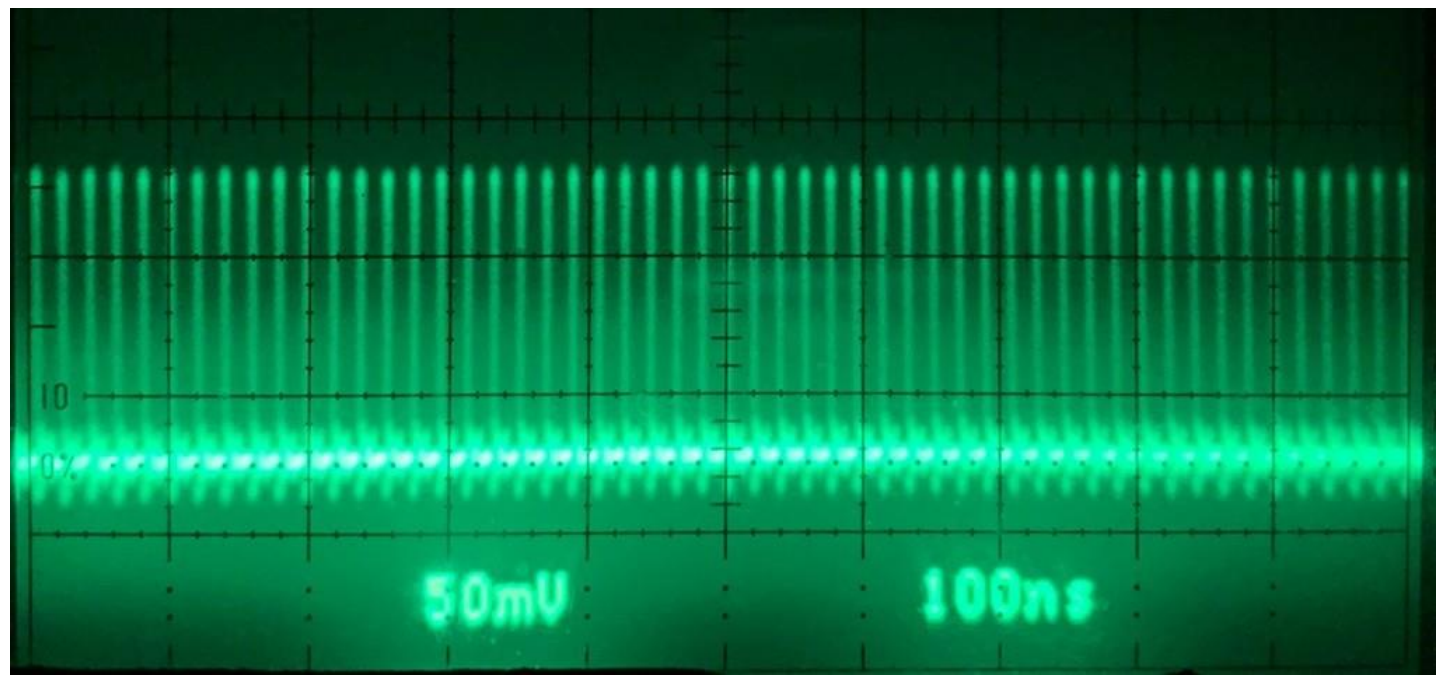

**(b)**

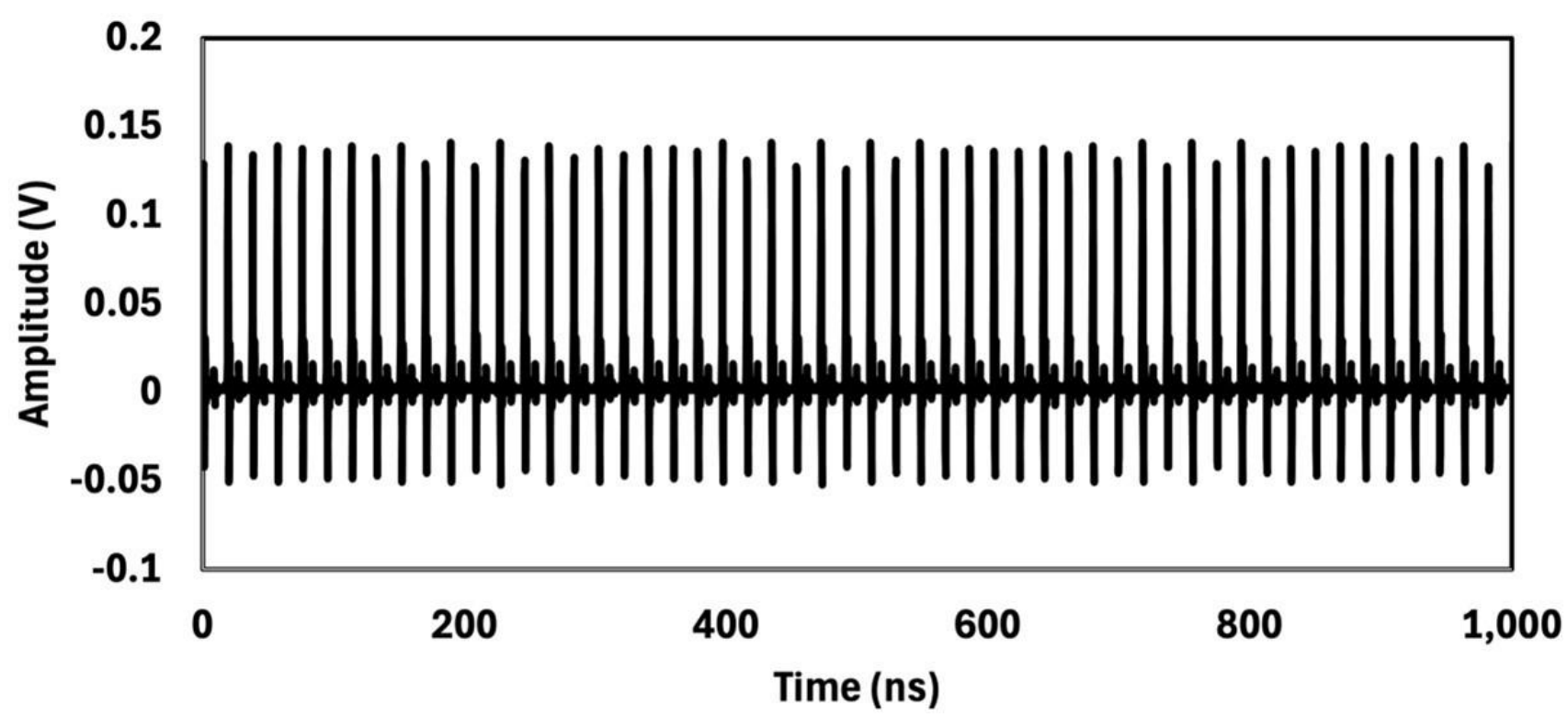

**(c)**

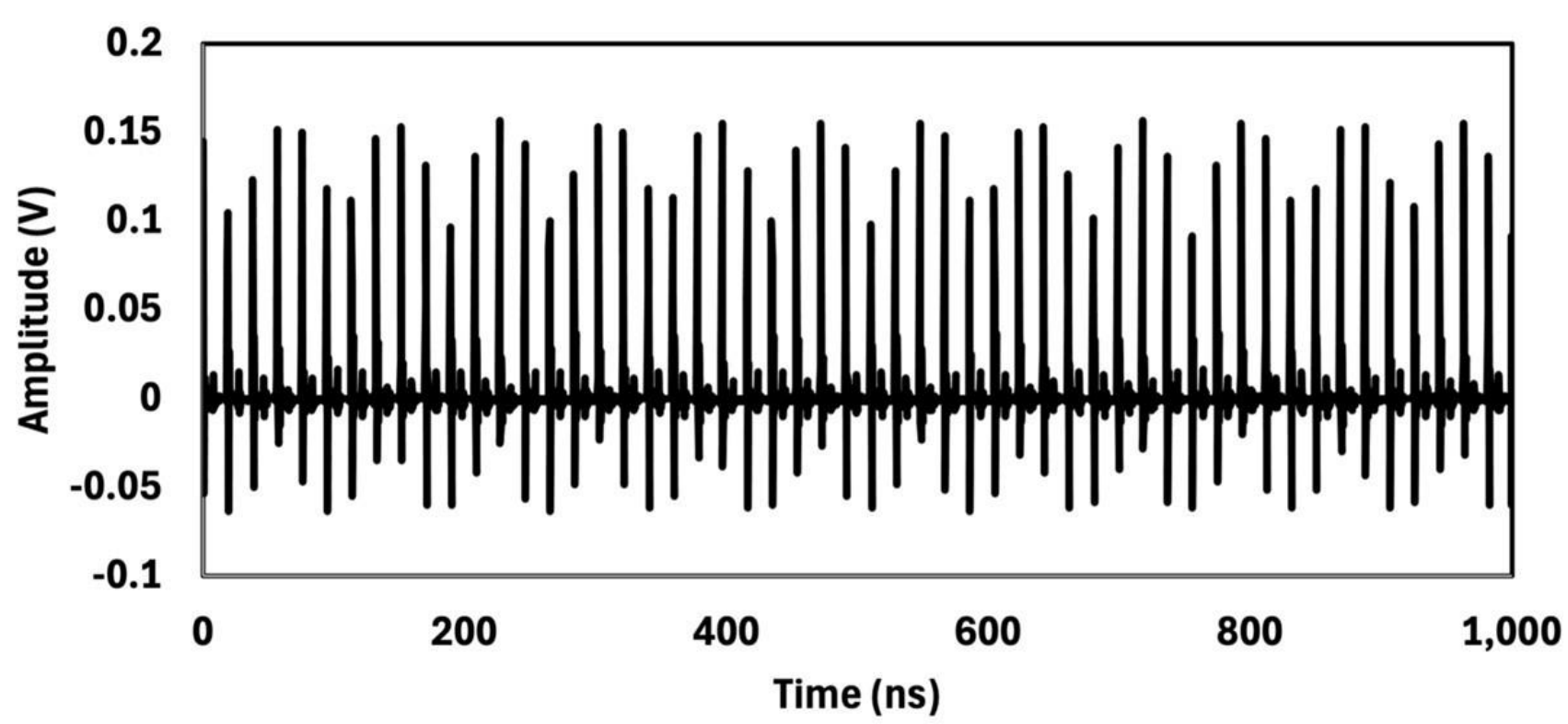

Fig. 1. Aliasing in repetitive signal acquisition. (a) Reference trace from an analog oscilloscope, showing the expected flat pulse-train envelope. (b) The same signal recorded with a digital oscilloscope at its full sampling rate, already showing deviations from the flat envelope due to sampling limitations. (c) The same signal recorded at half the maximum sampling rate, a common condition when multiple channels share the digitizer, where aliasing artifacts are further amplified into a pronounced beat-like modulation.

In the next sections we discuss the source of sampling error, detail the APF method, presenting its natural and unique advantages. We then give empirical illustrations of the method, applied to bare-photodiode measurements, light collection using an integrating sphere, and multimode-fiber collection coupled with a photodiode. Fitting errors are quantified and compared with those from spline fitting and sinc(x) interpolation. The pulse-archetype fitting method shows broad adaptability for higher-precision data collection at the limits of use of digitizing oscilloscopes.

## 2. Sources of sampling and limitations of conventional fitting

A survey of manufacturers shows oscilloscope design typically provides sampling rates 5 to 8 times the bandwidth of the pre-amplifiers6, which translates to about 4-6 samples over a pulse having that bandwidth. Taking, e.g., a Gaussian pulse sampled this way, the error in peak-height measurement due to mis-sampling the exact peak ranges roughly 6–13% (see Fig. 2).

Often – in the case of the measurement of ultrafast laser pulses, for example – it is the oscilloscope pre-amplifier that limits resolution of the pulse-shape. For very short electrical signals, the oscilloscope signal is an instrument-response function. Until the oscilloscope bandwidth approaches the signal pulse-bandwidth, typical sampling-rate design means that nothing much is gained in pulse-measurement by using a better oscilloscope. One simple improvement is to use smooth interpolation to partially reconstruct the missed peak. Spline fitting, or sinc(x) interpolation7, is often employed.

The method of sinc(x) interpolation would be a complete answer, if sampling met the Nyquist criterion. However, oscilloscope pre-amplifiers are not bandlimited and are measured in practice as full width at 70% of peak gain. Thus, Nyquist condition is not typically met, and sinc(x) interpolation, like spline-fitting, may still be constructive but is incomplete. Other general fitting methods may be helpful where they, in essence, extract consistency-information from local sample points around the peak. They may be unhelpful, when they are sensitive to data noise, which may cause over-fitting8.

In chief, none of these fitting methods takes advantage of information in the known pulse archetype, and therefore the fixed relationship between the pulse peak and the several measured data points scattered around it. With this uniform constraint, peak heights and timings can be found from conventionally undersampled data.

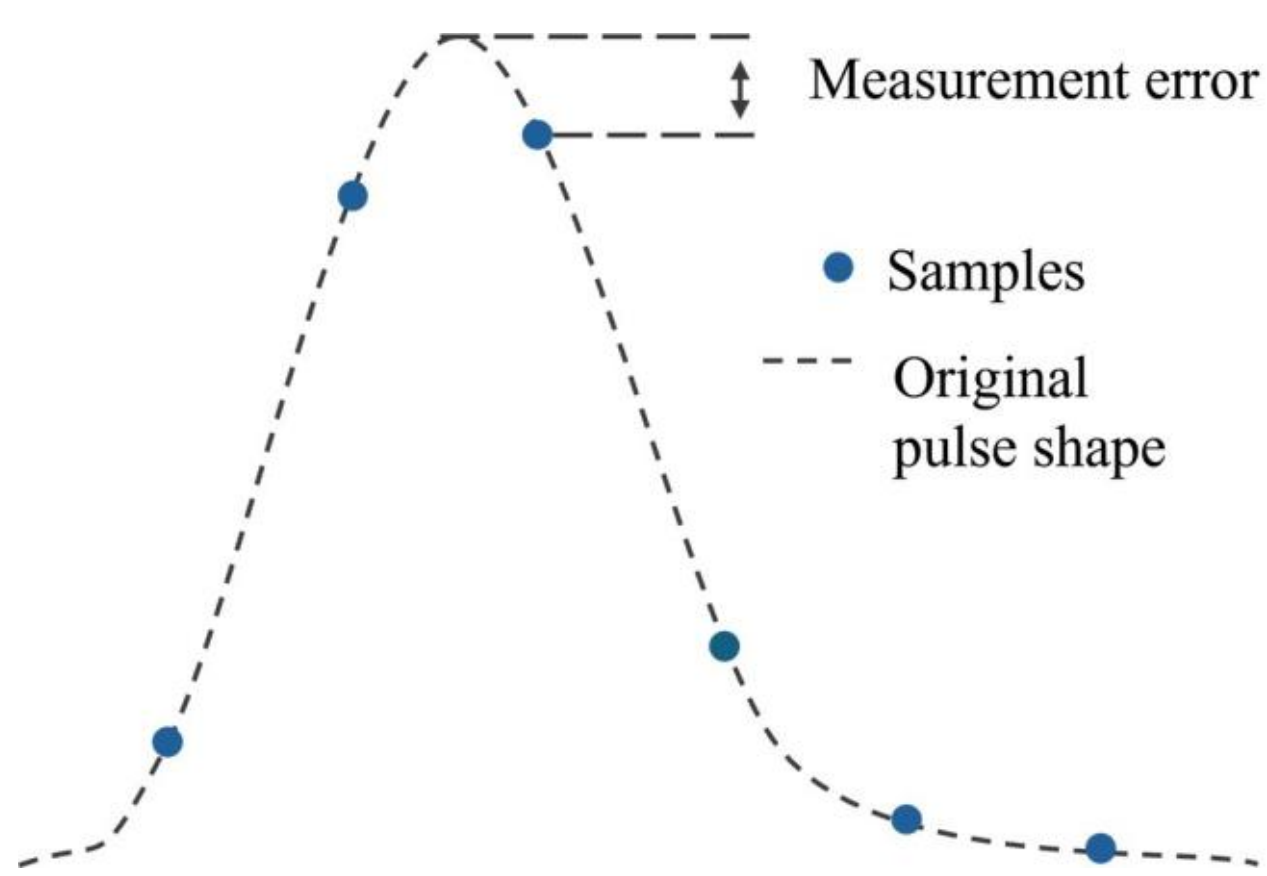

Fig. 2. Schematically illustrates the measurement error when sampling misses the exact pulse peak.: Typical oscilloscopes with sampling rates 5 to 8 times the analog bandwidth allows only 4-6 samples over an associated impulse-response function. In the case of a Gaussian pulse the error due to mis-sampling the exact peak ranges between 6% and 13%.

## 3. Principle and models of archetype parametric fitting

Our approach, archetype parametric fitting (APF), applies for any waveforms where all pulses have the same form, such as photodiode pulses generated from a mode-locked laser. Particularly, the approach works for any signals that are significantly shorter than the detecting system's response time, that is, for all signals so short that the detector or oscilloscope produces impulse-response functions. Traces of this kind retain just two simple pieces of information: the pulse amplitude (or rather the pulse energy), and the arrival time of the pulse.

Since the waveform is standard, we simplify the requirement of fitting arbitrary digitized waveforms to fitting a building-block, the pulse-shape archetype. The approach for each pulse reduces to finding best-fit parameters for pulse arrival time, amplitude and any potential vertical-bias offset (backgrounds). The method in principle even supports good characterization of pulses that have been digitized at sampling rates below the nominal Nyquist frequency, and without risk of overfitting.

For an example case of our experiments conducted, we approximate each pulse as a half-Gaussian for the rise, and an exponential tail, $f(t)$:

$$f(t) = \begin{cases} a_1 e^{-\frac{t^2}{2c^2}}, & t \leq \Delta t \\ a_2 e^{-\lambda(t-\Delta t)}, & t > \Delta t \end{cases} \tag{1}$$

where

$$a_2 = a_1 e^{-\frac{\Delta t^2}{2c^2}} \tag{2}$$

for continuity.

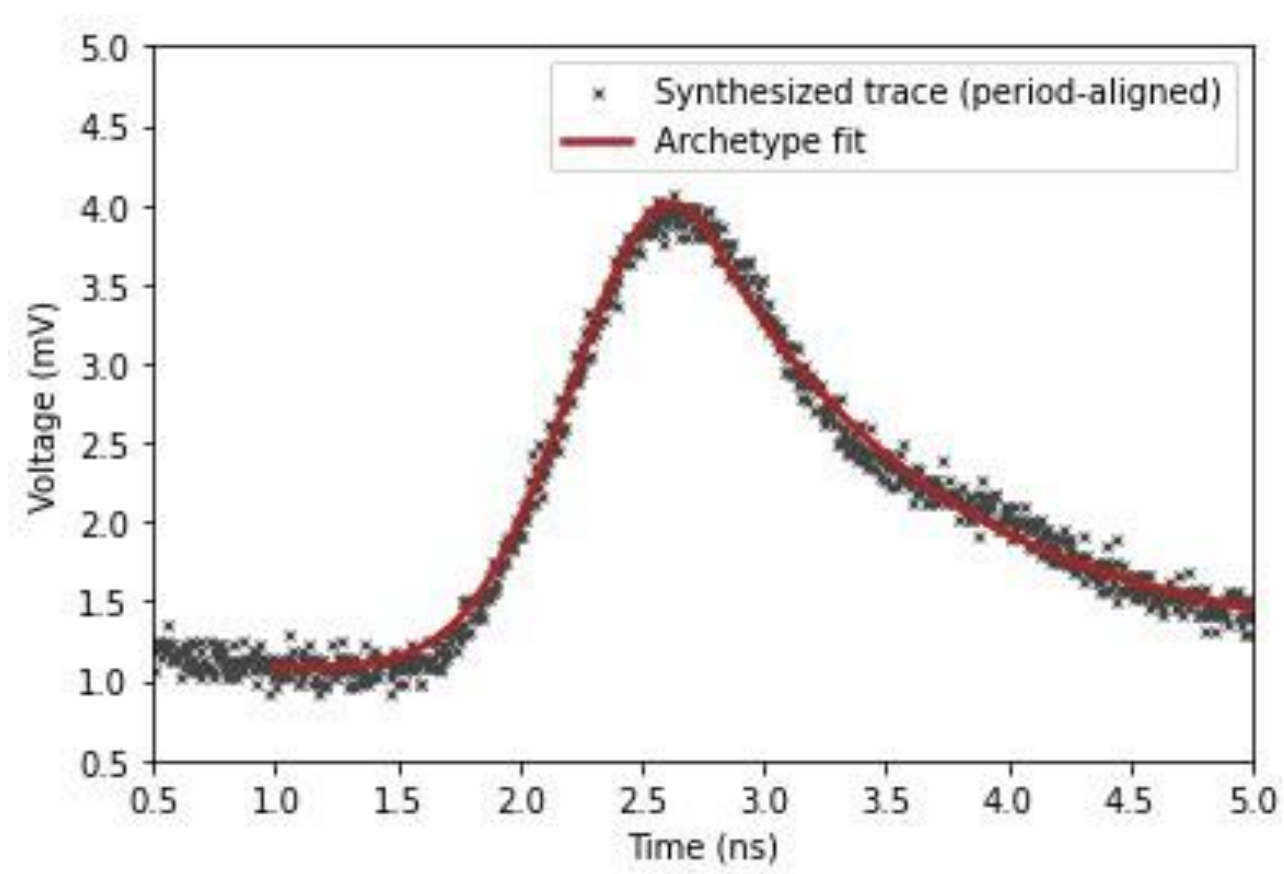

Fig. 3. Synthesized sampling-scope trace of a 350 fs laser pulse. The waveform was obtained by recording a stable 200 MHz pulse train with a photodiode (Thorlabs DET-10A2) and digital oscilloscope (LeCroy Wavesurfer 3054), then remapping successive pulses onto a single period (modulo the measured repetition rate). This procedure yields a densely sampled trace of the pulse-shape, serving as the archetype for subsequent fitting.

Δt sets the time to transition from gaussian to exponential. the fitting parameters $a_1$, $a_2$, $c$, and $\lambda$ can be characterized in detail by sampling scope-measurements directly, or synthesized artificially from one trace of a multi-pulse train as we show in Fig. 3. This wavefom was obtained from a stable 200 MHz pulse train provided by the laser's built-in diagnostic output. A sequence of pulses was remapped onto a single period (modulo the precisely measured repetition rate), producing a densely sampled trace analogous to a sampling oscilloscope.The precise knowledge of this base-element for all measurements represents the added information content that subsequently allows us to infer individual pulse amplitudes $c_n$, pulse timing $\tau_n$, and any trailing offset for the pulses, $h_n$ (see second paragraph below), within a modulated pulsetrain, from just a few sampled points across each pulse.

In the case of basic modelocked-laser pulsetrains, $\tau_n$, are uniform; in general the parameter $\tau_n$ can accomodate pulse reflections, or interdigitated or multiplexed pulsetrains[9,10] (see Fig. 4).

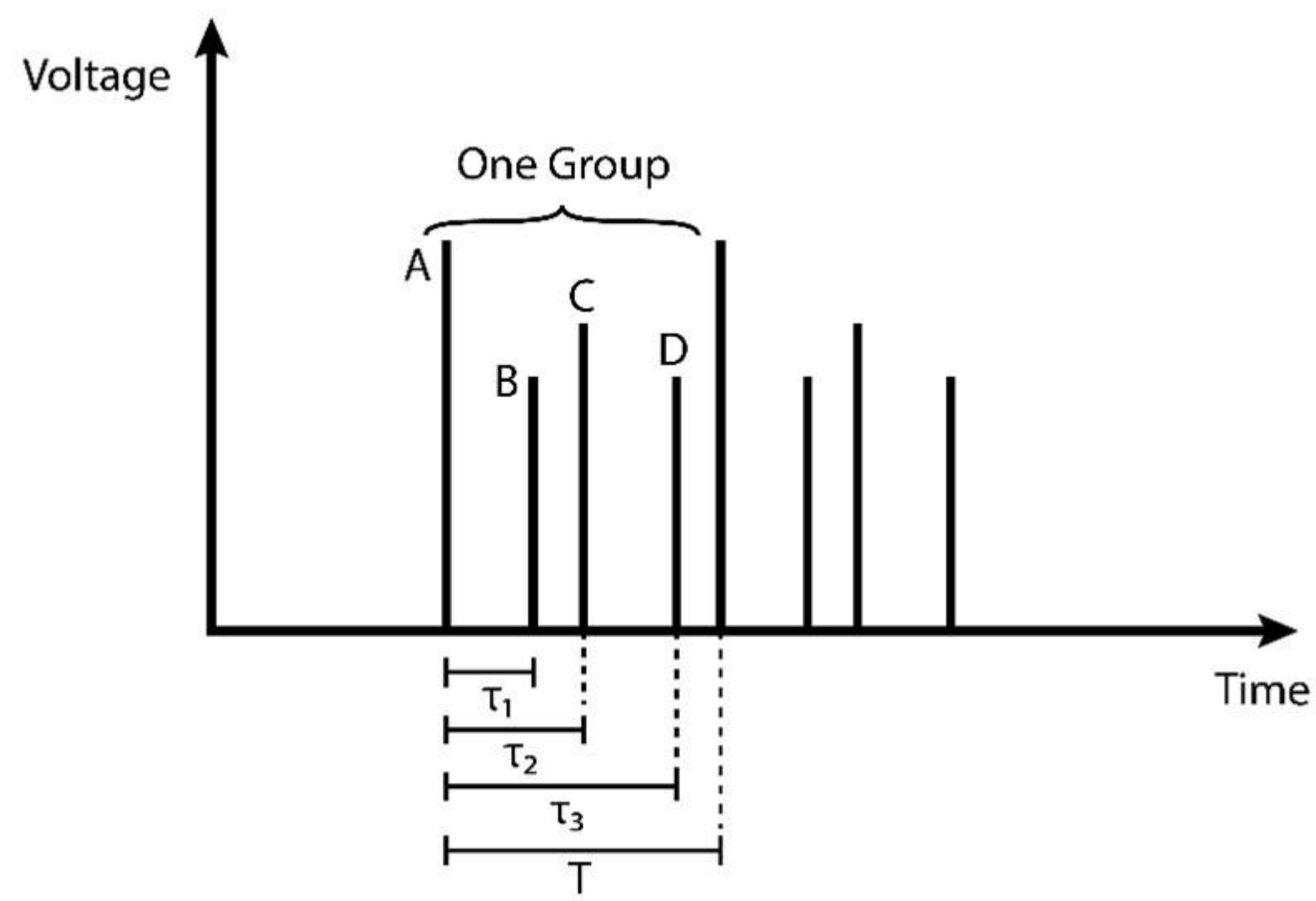

Fig. 4. Schematics of use of $\tau_n$ to describe a sub-cycle of 4 pulses in a train of base-period of T

Fitting whole pulsetrains, instead of pulse-by-pulse, also addresses a particular issue for pulsetrain measurements, caused by the risetime/falltime of the pulses within the waveform. When high-repetition rate signals have a pulse-repetition period only a little larger than the individual pulse durations, the rising edge of the next pulse waveform will sit on the exponentially decaying tail of the previous pulse waveform, the amplitude of which varies with the previous pulse amplitude. Unless the pulse rise is a step-function, it can be unreliable to estimate the pulse amplitude as the increment from trough to peak, because this background decays, from the time of trough to time of peak.

## 4. Application and evaluation

We have recently applied the APF method to evaluate its robustness across different detection systems in time-resolved laser–plasma reflectometry experiments[10,11] . In this test, a single pulse-train was sampled simultaneously by two integrating-sphere setups, each coupled to a different photodiode–oscilloscope chain. To further challenge the method, one of the traces was intentionally recorded at half the nominal sampling frequency, emulating an under-sampled detection condition (see Fig. 5). Despite these differences in detector models and sampling rates, the fitted pulse-amplitudes from APF analysis remained highly consistent: the coefficient of variation of amplitude-ratios between the two systems for a given burst was only ~2%.

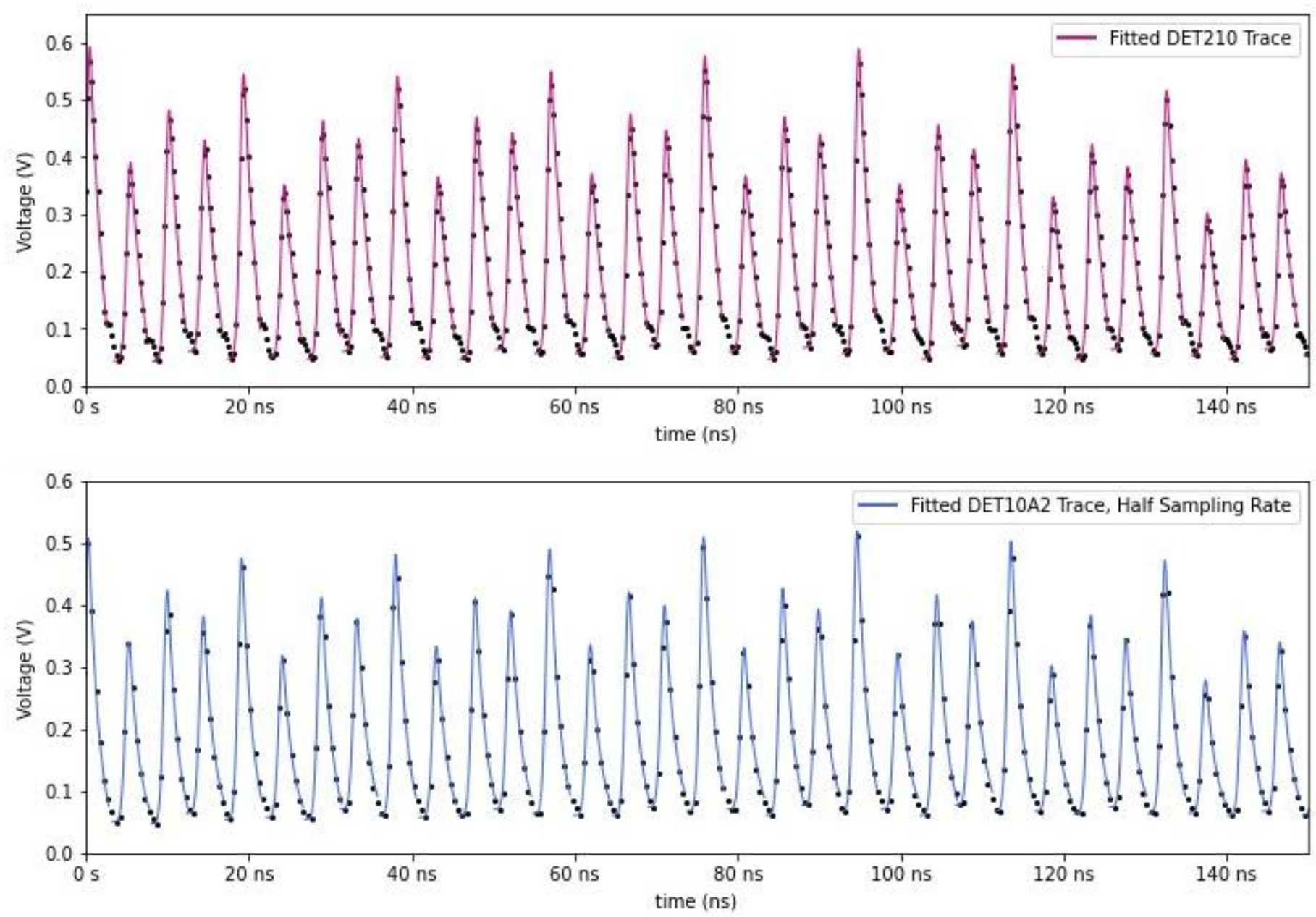

Fig. 5. . Fitted traces by the APF method for the same burst collected with different detection chains. Top: Reference trace acquired with an integrating sphere plus Thorlab DET210, digitized by a WaveSurfer 3054 oscilloscope. Bottom: Trace acquired with an integrating sphere plus Thorlab DET10A2 on another WaveSurfer 3054, recorded at half the nominal sampling frequency to emulate under-sampling conditions.

To quantify the APF fitting errors and to compare with existing more general methods, we generated data for pulses based on the mathematical function above, with pre-determined amplitudes, and then processed them using the APF method, cubic spline fit, and sinc(t) interpolation, at several sampling rates (see Fig. 6).

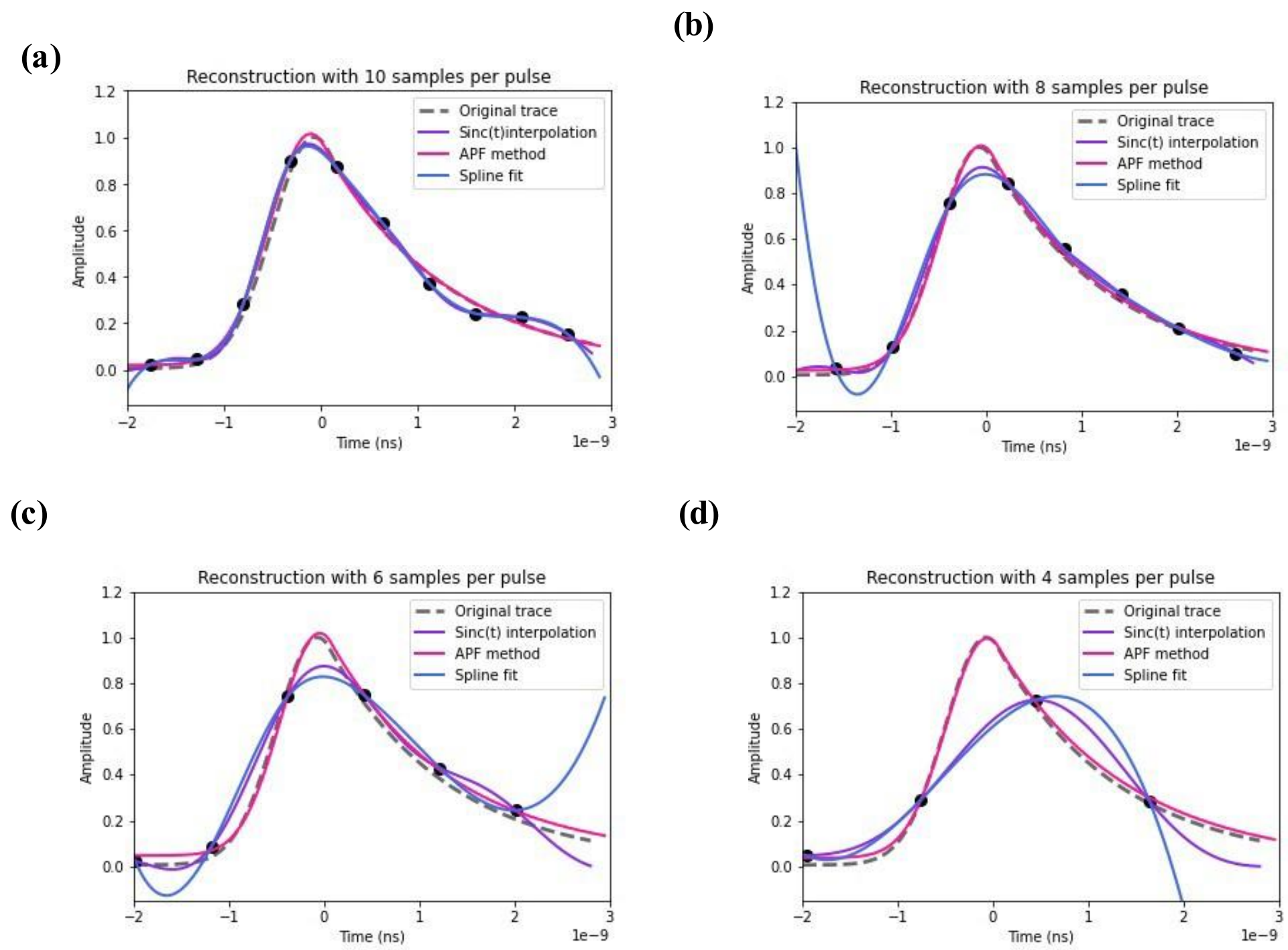

Fig. 6. . Fitting examples of APF method, compared with cubic spline fit and sinc(t) interpolation under different sample rates. Data points for artificial pulses were generated based on half Gaussian and half exponential model. Samples generated from synthetic pulses were subject to random fluctuations in amplitudes ranging from 0 to 0.05 in order to simulate digitizing errors and other noises found in applications. A random shift in time axis to all samples was added to simulate asynchronous sampling. The APF archetype pulse parameters were determined using a high sample rate trace produced by collapsing a train of 25 such pulses (modulo T), as described in text.

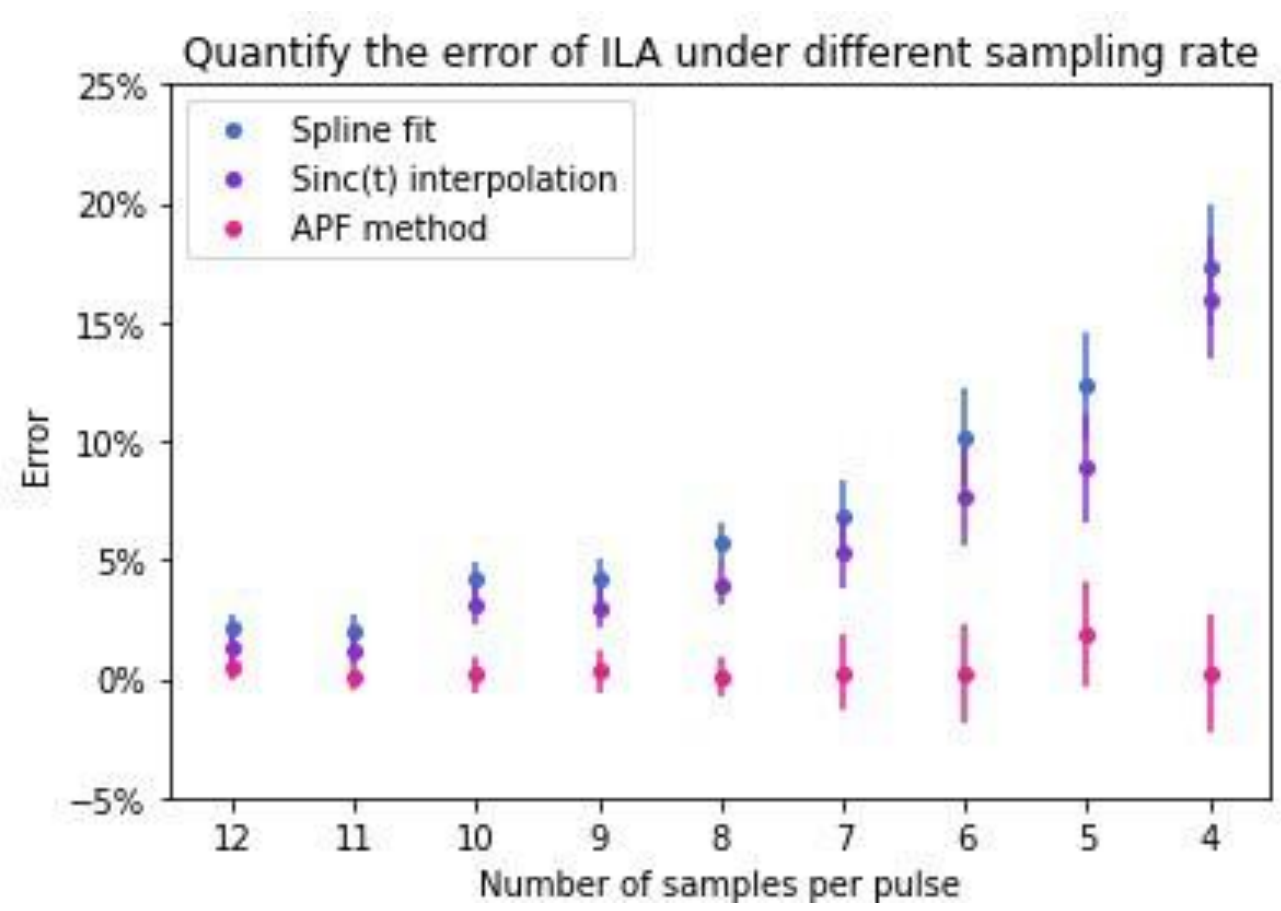

Fig. 7. Comparison of errors: APF vs. other fitting methods for a known pulse at progressively reduced sampling rates. (Averages of 20 trials with asynchronous sampling)

As Figs. 6 and 7 illustrate, for ample sampling rates ( 12 data-samples over a pulse) the different fitting approaches work equally well. As the sampling rate decreases to typical oscilloscope sampling rates (4-6 samples over a pre-amplifier bandwidth-limited pulse) and below, sinc(t) interpolation and spline fitting each exhibit growing fitting errors, whereas APF being constrained to the known pulse-shape, retains excellent agreement with the original artificial pulse, as shown in Fig. 6(d). Uncertainties do increase for the APF method with lower sampling rates. However, unlike cubic spline interpolation and sinc(x) interpolation, the fitted average values still give accurate determination of peak amplitude. This is an important strength of the method in undersampled cases: by using the known pulse shape, data samples a little distance away from the peak still give information about the whole pulse, which consists of the peak amplitude and relative timing.

## 5. CONCLUSION

We propose a specialized-pulse fitting method for sparsely sampled quasi-repetitive waveforms, especially for instrument-limited waveforms generated from ultrashort laser pulse-trains. Waveforms of this kind can be very profitably fitted by exploiting the known, detailed, pulse-archetype as a graphical basis vector. Where other more-general fitting methods use adjacent sampled-data values to “fill in” the waveform by general interpolation, we take advantage of the knowledge of what each pulse shape necessarily must be, and find pulse arrival times, amplitudes and vertical-bias offsets (backgrounds). This makes the APF method a particularly profitable strategy even at sampling rates below the ostensible Nyquist frequency. Key factors that allow this are knowledge of detector- or instrument-limited pulse shapes: a repetitive pulse shape can be known to far greater precision than nominal sampling rates afford, and this additional information supplements the fitting done from sparse sampling.

By knowing the generic pulse shape very well, and by fitting the entire pulse train simultaneously, the method meaningfully identifies pulse amplitudes even in the context of a decaying tail from previous pulses (pulse partial overlaps) where other methods are not well-disposed to interpret. We have demonstrated the APF strategy in detailed time-resolved reflectometry measurements, showing its broad adaptability and its consistency across different detection systems. Error analysis demonstrates that the APF method significantly improves quantitative measurement, compared to other more general fitting methods, especially under conditions of undersampling.

## ACKNOWLEDGMENTS

We acknowledge the support of the Natural Sciences and Engineering Research Council of Canada (NSERC). RGPIN-2017-06757

Nous remercions le Conseil de recherches en sciences naturelles et en génie du Canada (CRSNG) de son soutien. RGPIN-2017-06757

We acknowledge University of Toronto Arts Science Tri-Council Bridge Funding support, through the Vice Dean, Research & Infrastructure, Faculty of Arts and Science.

## Author Declarations

### *Conflict of Interest Statement*

The authors have no conflicts to disclose.

### *Author Contributions*

Jingsen Tang: Formal analysis (lead); Funding acquisition (supporting); Investigation (lead); Software (supporting); Supervision (supporting); Validation (lead); Visualization (lead); Writing – original draft (lead); Writing – review & editing (equal). Nicolas Souleles: Conceptualization (supporting); Formal analysis (supporting); Investigation (supporting); Methodology (equal); Software (equal); Validation (supporting); Writing – original draft (supporting). Anna Hwang: Conceptualization (supporting); Formal analysis (supporting); Investigation (supporting); Methodology (equal); Software (equal); Validation (supporting). Luke Coulter: Investigation (supporting); Writing – review & editing (supporting). Denisa Bani: Formal analysis (supporting); Investigation (supporting). Robin S. Marjoribanks: Conceptualization (lead); Funding acquisition (lead); Resources (lead); Methodology (supporting); Supervision (lead); Writing – original draft (supporting); Writing – review & editing (equal).

## DATA AVAILABILITY

The data that support the findings of this study are available from the corresponding author upon reasonable request. Example implementation of the code is available in GitHub at (https://github.com/NicoSouleles/Burst-Fit-V2)

## REFERENCES

[1]Qian, Z., Covarrubias, A., Grindal, A. W., Akens, M. K., Lilge, L., & Marjoribanks, R. S. (2016). Dynamic absorption and scattering of water and hydrogel during high-repetition-rate (> 100 MHz) burst-mode ultrafast-pulse laser ablation. Biomedical Optics Express, 7(6), 2331-2341.

[2]Chowdhury, I. H., Wu, A. Q., Xu, X., & Weiner, A. M. (2005). Ultra-fast laser absorption and ablation dynamics in wide-band-gap dielectrics. Applied Physics A, 81, 1627-1632.

[3]Puerto, D., Gawelda, W., Siegel, J., Bonse, J., Bachelier, G., & Solis, J. (2008). Transient reflectivity and transmission changes during plasma formation and ablation in fused silica induced by femtosecond laser pulses. Applied Physics A, 92, 803-808.

[4]Metzner, D., Lickschat, P., Kreisel, C., Lampke, T., & Weißmantel, S. (2022). Study on laser ablation of glass using MHz-to-GHz burst pulses. Applied Physics A, 128(8), 637.

[5]Incoronato, A.; Cusini, I.; Pasquinelli, K.; Zappa, F. Single-shot pulsed-LiDAR SPAD sensor with on-chip peak detection for background rejection. IEEE J. Sel. Top. Quantum Electron. 2022, 28, 1–10.

[6]Tektronix. (n.d.). Evaluating Oscilloscopes: Learn About Key Features & Functions. https://www.tek.com/en/documents/primer/evaluating-oscilloscopes

[7]Crochiere, R. E., & Rabiner, L. R. (1983). Multirate digital signal processing (Vol. 18). Englewood Cliffs, NJ: Prentice-Hall.

[8]Isaacson, E., & Keller, H. B. (1994). Analysis of numerical methods. Courier Corporation.

[9]Kerse, C., Kalaycıoğlu, H., Elahi, P., Akçaalan, Ö., & Ilday, F. Ö. (2016). 3.5-GHz intra-burst repetition rate ultrafast Yb-doped fiber laser. Optics Communications, 366, 404-409.

[10]Haboucha, A., Zhang, W., Li, T., Lours, M., Luiten, A. N., Le Coq, Y., & Santarelli, G. (2011). Optical-fiber pulse rate multiplier for ultralow phase-noise signal generation. Optics letters, 36(18), 3654-3656.

[11]Marjoribanks, R. S., Tang, J., Dzelzainis, T., Prickaerts, M., Lilge, L., Akens, M., Veevers, C., Gharabaghi, N. N., Hitzler, A., Kalaycioglu, H., Yavas, S., & Karamuk, S. G. (2024). Ultrashort-pulse burst-mode materials processing and laser surgery.. In D. Yang (Ed.), Pulsed laser processing of materials (pp. 111–137). IntechOpen. doi: 10.5772/intechopen.1005152

[12]Marjoribanks, R. S., Tang, J., Dzelzainis, T., Prickaerts, M., Lilge, L., Akens, M., ... & Ilday, F. O. (2024, March). Plasma persistence, accumulated absorption, and scattering: what physics lets us control the heat left behind in ultrafast-pulse burst-mode laser surgery. In Frontiers in Ultrafast Optics: Biomedical, Scientific, and Industrial Applications XXIV (Vol. 12875, pp. 9-19). SPIE.